\documentclass[%
 reprint,
 amsmath,amssymb,
 aps,
prl,
]{revtex4-2}

\usepackage{graphicx}
\usepackage{dcolumn}
\usepackage{bm}

\newcommand{\rem}[1]{ } 
\newcommand{\beq}{\begin{equation}}
\newcommand{\eeq}{\end{equation}}

\newcommand{\aap}{{Astron. \& Astroph.}}
\newcommand{\araa}{{Ann. Rev. Astron. Astroph.}}
\newcommand{\mnras}{{Mon. Not. R. Astron. Soc.}}
\newcommand{\apjl}{{Astrophys. J. Lett.}}

\begin{document}


\title{Origin of spectral bands in the Crab pulsar radio emission}

\author{Mikhail V. Medvedev}
\affiliation{Institute for Advanced Study, School for Natural Sciences, Princeton, NJ 08540}
\affiliation{Department of Astrophysical Sciences, Princeton University, Princeton, NJ 08544}
\affiliation{Institute for Theory and Computation, Harvard University, Cambridge, MA 02138}
\affiliation{Department of Physics and Astronomy, University of Kansas, Lawrence, KS 66045}
\affiliation{Laboratory for Nuclear Science, Massachusetts Institute of Technology, Cambridge, MA 02139}


\begin{abstract}
The model explaining the spectral ``zebra'' pattern of the high-frequency interpulse (HFIP) of the Crab pulsar radio emission is proposed. The observed emission bands are diffraction fringes in the spectral domain. The pulsar's own plasma-filled magnetosphere plays a role of a frequency-dependent ``diffraction screen''. The observed features such as the proportional band spacing, high polarization, constant position angle, and others are explained. The model allows one to perform ``tomography'' of the pulsar magnetosphere. Indeed, we have obtained the plasma density profile directly from observations, without assuming a particular magnetosphere. Our model is testable and several predictions are made. The two ``high-frequency components'' observed at the same frequencies as the HFIP are proposed to be related to HFIP. 
\end{abstract}

\maketitle


\section{Introduction}

Pulsars are spinning magnetized neutron stars. Pulsar properties and theories are reviewed in Ref. \citep{PK22}. Most pulsars exhibit one or, rarely, two radio pulses per rotation period. Radio emission is believed to be produced in the inner magnetosphere in the polar cap region \citep{P+20}. Thus, it appears in a different rotation phase than the high-energy emission, which is thought to be produced outside the ``light cylinder'' (LC). (The LC is defined as the region where linear speed approaches the speed of light, $\Omega\, r_{LC}\sim c$, where $\Omega$ is the pulsar rotation angular velocity.) The Crab pulsar is, in contrast, very special. Its radio main pulse and interpulse are coincident in phase with high-energy emission, indicating the same emission region. It is now believed that the emission occurs in the {\em caustic} regions in the current sheet, near the LC \citep{BS10}. Broadband spectrum is the general feature of the main pulse, low-frequency interpulse, and several emission components in the frequency range between about 1~ GHz and 43~GHz studied in Refs. \citep{HE07, HE+16b}.

The spectral pattern of the {\em high-frequency interpulse} (HFIP), observed between about $\nu\sim5$~GHz and $\nu\sim30$~GHz, is remarkably different and represents a sequence of emission bands resembling the ``zebra'' pattern. This peculiar spectral patten was first reported in 2007 and subsequently studied in great detail \citep{HE07, HE+16b}. 

Here are the known properties of the HFIP emission any successful model should account for. 
\begin{itemize}
\item The very existence of peculiar spectral features in the form of regular emission bands.
\item The band proportional separation and the ``6\%-rule''. The frequency difference between the nearest bands is proportional to the band frequency, so that $\Delta\nu\approx0.057\nu$.
\item The pattern persistence. There has not been observed a single HFIP without spectral bands.
\item The pattern stability. The band positions can be stable for as long as a day, though they can also vary from a pulse to a pulse.
\item The HFIP is nearly 100\% linearly polarized and the position angle is stable and does not change over many pulses.
\item The HFIP has a variable and often larger dispersion measure (DM) than the main pulse. 
\end{itemize}


Despite strong and extensive theoretical efforts, no satisfactory mechanism explaining the HFIP puzzle has been proposed for over fifteen years. The proposed models either involve emission mechanisms producing multiple harmonics (e.g., cyclotron or maser emissions) , or invoke propagation effects (e.g., interference at a source, within a current sheet, or the wave modulation instability). The former class of models predicts the incorrect uniform band spacing. The latter class requires very high wave phase coherence and source stability, which are unreasonable to expect in a highly turbulent circum-pulsar medium. The detailed review of available models can be found in Ref. \citep{HE16c,Benachek+24}. In this paper, we propose a robust model that explains the HFIP puzzling properties and makes testable predictions.

\section{Model}

\begin{figure}
\includegraphics[scale=0.22]{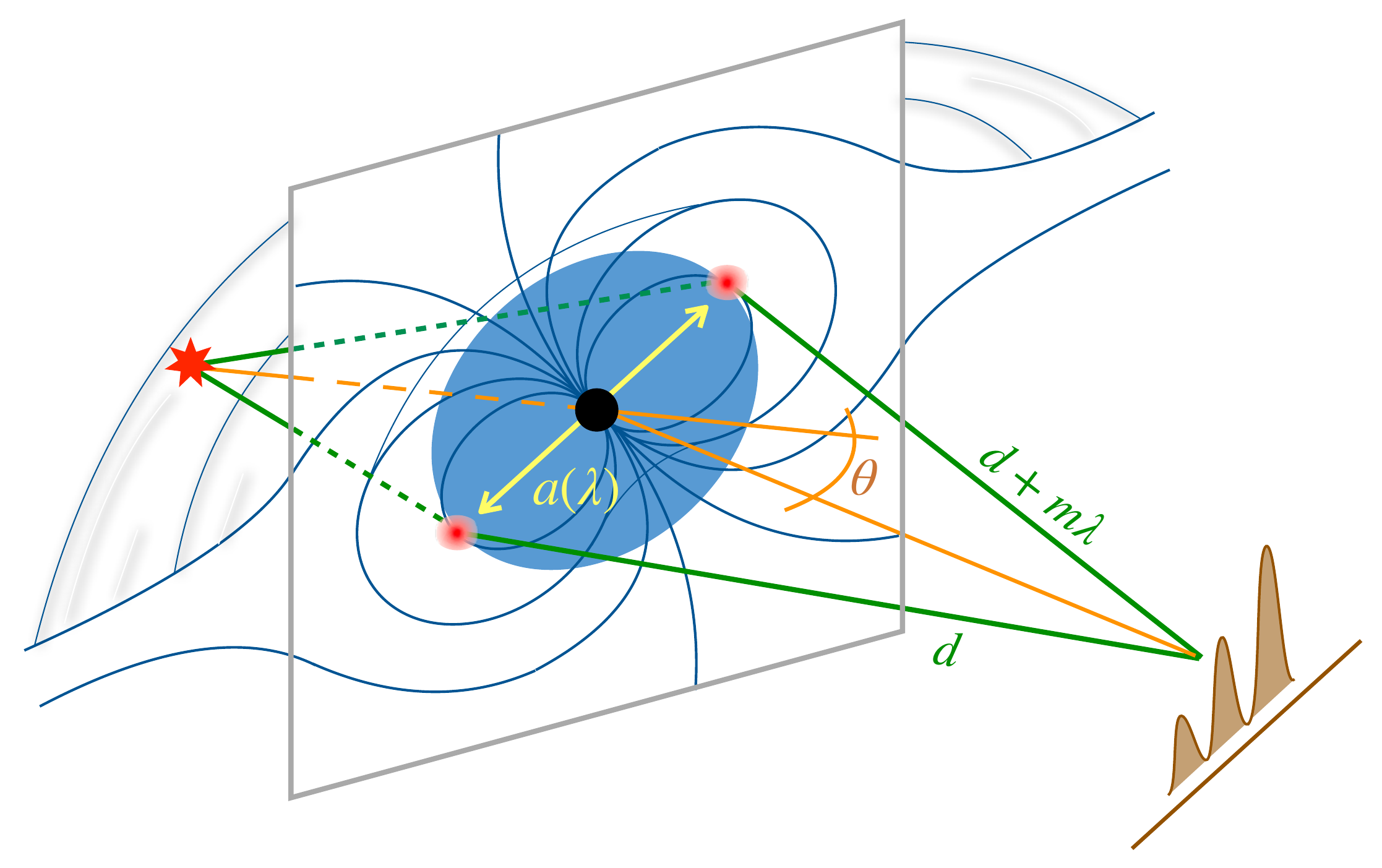}
\caption{Overall geometry of the system. }
\label{f:pulsar}
\end{figure}

We propose that the observed spectral band structure is the diffraction fringe pattern off the pulsar magnetosphere. Here are the postulates of the model. 

First, we postulate that the broadband radio source is located behind the neutron star, as in Fig \ref{f:pulsar}. This is the main assumption. We do not need to specify the origin of the source. However, it is natural to assume, that the mechanism is the same as in the main pulse and the low-frequency interpulse \citep{P+19}, see Appendix for dicussion. Alternatively, a reflected or refracted radio emission from elsewhere (even a polar cap) can serve as the needed source. 

Second, we assume that the extraordinary (X-mode) is strongly cyclotron-absorbed, so that only the ordinary (O-mode) can propagate through the magnetosphere and reach the observer, see Appendix for details. This consideration explains the nearly 100\% polarization of the HFIP. The constancy of the position angle is explained by pure geometry: the radio source is always located behind the magnetosphere at the same rotation phase. Variations in the position angle and circular polarization, if any, can be affected as plasma normal modes propagate through the dynamical magnetospheric plasma \citep{PL00,BP12}, as well as variations at the source.

Third, the magnetosphere is filled with plasma whose density is expected to decrease with distance. We consider closed fields lines only since this part of the magnetosphere will obscure the open field lines in the region of interest anyway. 
O-mode has the refraction index \citep{M23}:
\begin{equation}
n_{\rm refr}^{2}\equiv\left({kc}/{\omega}\right)^2=1-{\omega_p^2}/{\omega^2}.
\label{nOmode}
\end{equation}
where $\omega_{p}=\sqrt{4 \pi e^{2} n_{e}/\langle\gamma^{3}\rangle m_{e}}$ is the relativistic plasma frequency with $n_{e}$ being the electron (and positron, if any) density and $\langle\dots\rangle$ denoting an ensemble-averaged quantity \footnote{Plasma can be highly relativistic, $\gamma\gg1$, on open field lines and sub-relativistic on closed field lines, because of cooling of the trapped plasma.}. Since we assume that diffraction happens in the inner magnetosphere on closed field lines, one expects $\gamma\sim 1$. 
Close to the surface, a low-$\omega$ radio wave cannot propagate. Above a certain `reflection' radius, $r_{0}$, the index of refraction becomes positive and it can propagate. The reflection condition, $\omega=\omega_{p}(r)$, readily follows from the vanishing index of refraction, $n_{\rm refr}=0$. 

Thus, (i) only the O-mode can propagate and diffract, hence the observed emission is highly polarized, (ii) the ``diffraction screen'' is frequency dependent, and (iii) the magnetic field is nearly orthogonal to the line of sight near the ``screen'' (otherwise, the wave becomes oblique and its cyclotron absorption increases). Note that this geometry (i.e., the back-lighting the magnetosphere) explains the larger variation in the dispersion measure of the HFIP (compared to the main pulse) as simply the longer optical path of the radio emission through a denser magnetospheric plasma. The estimate of the excess dispersion measure, $\delta{\rm DM}=n_{e}r_{\rm pc}\sim0.03\textrm{ pc cm}^{-3}$, is fully consistent with observed values, $\delta{\rm DM}\sim0.02-0.04\textrm{ pc cm}^{-3}$ \citep{HE+16b}. 
Here $r_{\rm pc}$ is the size in parsecs, and we assumed $n_{e}\sim10^{10}\textrm{ cm}^{-3}$ in the magnetosphere at a distance $r\sim10^{7}$~cm, i.e., at ten star radii (see below). This dispersion can also smear out nanoshots, which are seen in the main pulse and low-frequency interpulse, but not in the HFIP (assuming the same mechanism of radio emission generation for all the pulses). 

Finally, to simplify analytical calculations, (i) we assume that $n_{\rm refr}=1$ above the reflection radius, $r>r_{0}$, and (ii) we approximate the spherical ``screen'' $r<r_{0}$ as a thin two-dimensional screen orthogonal to the line of sight. We do not expect the 2D thin screen to be circular. It is more likely to have two (or more) ``hot spots'' (as in Fig. \ref{f:pulsar}) rather than a uniformly illuminated annulus, given the absence of spherical symmetry of the magnetosphere
\footnote{As is true for any diffraction pattern, the pattern is seen in the plane containing a source and two hot spots. In the case of multiple hot spots, there are many such diffraction planes. In the case of an ``hot spot'' annulus, one obtains the annular diffraction pattern.}.
The hot spots are expected be especially bright when the plasma dispersion (having the index of refraction $n_{\rm refr}<1$) leading to the divergence of light rays is compensated by gravitational lensing (gravitational redshift also affects propagation via $n_{\rm refr}(\omega)$). In this paper, however, we do not consider general relativistic effects any further, assuming that diffraction occurs at radii, where these effects are small. In this case, it is reasonable to assume the hot spot separation to be approximately  $2r_{0}$. Consideration of gravitational corrections will be presented in a forthcoming publication. Numerical modeling of a polarized wave propagation in a realistic pulsar magnetosphere with variable index of refraction is desirable in the future.

Now, let us assume, for concreteness, that $\nu_{1}=\omega_{1}/2\pi=c/\lambda_{1}$ is the lowest frequency of a spectral fringe observed. For the HFIP, it is $\nu_{1}\simeq5$~GHz. This fringe's `order' (i.e., its number) is $m_{1}$. It is a free parameter in the model. The next interference fringe at frequency $\nu_{2}$ should overlap with the first one at the same spatial position given by a certain value of the deflection angle $\theta=const$, see Fig. \ref{f:pulsar}. Thus, its order is either $m_{1}+1$ or $m_{1}-1$. The third $\nu_{3}$ has the order $m_{1}+2$ or $m_{1}-2$, and so on. The last observed fringe is at $\nu_{N}\simeq30$~GHz and $N\simeq30$, as follows directly from observations. 
Thus the fringe overlap conditions are
\begin{align}
a_{1}\sin\theta&=m_1\lambda_1, \label{a1}\\
a_{2}\sin\theta&=(m_1\pm1)\lambda_2, \label{a2}\\
&\dots\nonumber\\
a_{N}\sin\theta&=(m_1\pm(N-1))\lambda_N,\label{aN}
\end{align}
where $a(\omega)=2r_{0}(\omega)$ is the frequency-dependent size of the ``screen'', see Fig. \ref{f:pulsar}, and $a_{k}\equiv a(\omega_k)$ for $1\le k\le N$. The apparent screen size for $k$-th band is 
\begin{equation}
a_k=(m_1\pm(k-1))2\pi c/\omega_k\sin\theta.
\label{ak}
\end{equation}
The spectral bands are distributed proportionally, so that $\omega_{k+1}-\omega_{k}=\omega_{k}\delta$, where $\delta\approx0.057$ as follows from the data fit \citep{HE+16b}. Hence, $k$-th  band frequency is
\begin{equation}
\omega_k=(1+\delta)^{(k-1)}\omega_1.
\label{wk}
\end{equation}

Now, we readily obtain the exact result for the apparent size as a function of frequency:
\begin{align}
a(\omega)=a_1\frac{\omega_1}{\omega}\left(1\pm\frac{(k-1)}{m_1}\right)
=a_1\frac{\omega_1}{\omega}\left(1\pm\frac{\log_{1+\delta}(\omega/\omega_1)}{m_1}\right).
\label{a-exact}
\end{align}
Hereafter, we drop the subscript $k$ in $\omega_{k}$, for simplicity.
We first note that the maximum value of $(k-1)/m_{1}$ does not exceed $N/m_{1}$. Next, if the total number of fringes produced in space is much larger than the number of observed ones, $N_{\rm tot}\gg N$, then the order of any observed fringe, $m$, should be substantially larger as well, since, statistically $m\sim N_{\rm tot}/2$. This simply states that there is a larger chance to observe higher-order fringes from the ``bulk'' of the pattern than those near its edge $m\lesssim N$. Furthermore, if $N/m_{1}\ll1$, the expression in the brackets represents the first two terms of the Taylor expansion of an exponential function, and also $\log(1+\delta)\simeq \delta$. Combining these altogether, we obtain:
\begin{equation}
1\pm\frac{\log_{1+\delta}(\omega/\omega_1)}{m_1}
\approx\exp\left(\pm\frac{\log(\omega/\omega_1)}{m_1\delta}\right)
\approx\left(\frac{\omega}{\omega_{1}}\right)^{\pm\frac{1}{m_1\delta}}.
\end{equation}

Now, the apparent size becomes a power-law function:
\begin{equation}
a(\omega)=a_1\left(\frac{\omega}{\omega_1}\right)^{-\beta},\quad \beta=1\mp\frac{1}{m_1\delta}.
\label{aw}
\end{equation}
Since $a$ is determined as twice the reflection radius, where $\omega_{p}=\omega$, we have (omitting the subscript $0$ in $r_{0}$)
\begin{equation}
4\pi e^2 n_e(r)/\langle\gamma^{3}\rangle m_e=\omega^{2}(r).
\label{reflect2}
\end{equation}
Solving Eq. (\ref{aw}) for $\omega$ one has, $\omega/\omega_{1}=(r/r_{1})^{-1/\beta}$, where $r_{1}\equiv r_{0}(\omega_{1})=a_{1}/2$. Thus, we can recover the plasma density profile in the magnetosphere:
\begin{equation}
n_{e}(r)=n_{e}(r_{1})\left(\frac{r}{r_{1}}\right)^{-\alpha},\quad \alpha=\frac{2}{1\mp\frac{1}{m_1\delta}}\sim2.
\label{ne}
\end{equation}
{\em We emphasize that we have obtained the power-law density profile strictly from observations, without making any theoretical assumption about the profile shape. }

\begin{figure}
\includegraphics[scale=0.92]{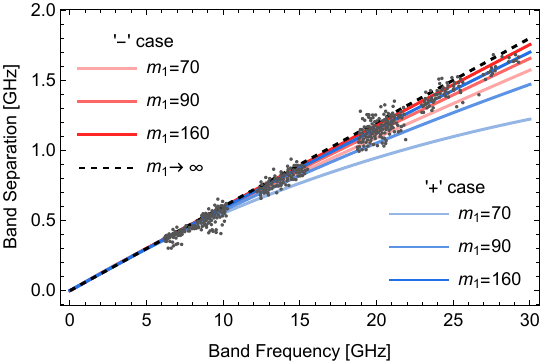}
\caption{Plot of $\Delta\nu$ vs. $\nu$ against approximate data points taken from \citep{HE+16b}, for several values of $m_{1}=70,\ 90,\ 160$ (light, medium, and dark colors, respectively) and $\delta=0.06$. Blue corresponds to ``$+$'' and red corresponds to ``$-$'' in Eqs. (\ref{aw},\ref{ne}). The dashed black line corresponds to $m_{1}\to\infty$, i.e., the exact law $\Delta\nu=\nu\delta$.
These curves correspond to $\alpha\simeq 1.6,\ 1.7,\ 1.8$ (blue) and $\alpha\simeq 2.6,\ 2.45,\ 2.2$ (red).}
\label{f:fit}
\end{figure}

The power-law density scaling has been obtained under the assumption that we are observing large-order fringes $m_{1}\gg N$. We now quantify this condition by comparing our result with the observed $\Delta\nu$ vs. $\nu$ correlation, as in Fig. \ref{f:fit}. Here, the curves obtained with Eq. (\ref{aw}) for $m_{1}=70,\ 90,\ 160, \ \infty$ are shown. Red and blue colors correspond to the upper and lower sign in Eq. (\ref{aw}), respectively. The by-eye inspection yields that a theoretical curve in-between the black dashed line and the nearby dark-blue curve would yield an acceptable fit to the data. These curves have $m_{1}\gtrsim110$ for the ``$-$'' case and $m_{1}\gtrsim160$ for the ``$+$'' case. With Eq. (\ref{ne}), these values translate into the following constraint:
\begin{equation}
n_e(r)\propto r^{-\alpha},\quad 1.8\lesssim\alpha\lesssim2.4.
\end{equation}
A true fit to the actual observational data is desirable.

\section{Discussion}


Let us recall that the Goldreigh-Julian \citep{GJ69} density is the minimum density needed to be present in the magnetosphere for it to co-rotate with the pulsar as a rigid body up to the light cylinder,
\begin{equation}
n_{GJ}(r)=\frac{\Omega B}{2ce}\simeq0.07\frac{B}{P}\textrm{ cm}^{-3},
\end{equation}
where $B$ is the magnetic field strength in gauss and $P=2\pi/\Omega$ is the spin period in seconds. The actual density of plasma in the magnetosphere can be larger than $n_{GJ}$ by a factor of multiplicity ${\cal M}>1$. For the Crab pulsar with the period of $P\simeq0.0335$~s and the magnetic field $B(r)\simeq4\times10^{12}(r/R_*)^{-n}$~gauss ($R_*$ being the neutron star radius), we have for the plasma density:
\begin{equation}
n_{e}(r)\simeq8.5 \times10^{12}\ {\cal M} \left(\frac{r}{R_*}\right)^{-n}\textrm{ cm}^{-3}.
\end{equation}
Here the index $n=2$ describes a monopolar (wind-like) configuration and $n=3$ corresponds to the dipolar field. Our results lie near 2 and slightly ``lean'' toward 3, assuming ${\cal M}={\rm const}$. The deviation of the index from 3, expected for a dipolar field, can, for example, be due to the radial dependence of ${\cal M}(r)$, the diffraction being in the polar regions, where $n_{e}\propto r^{-2}$ in the pulsar wind. 

\begin{figure}
\includegraphics[scale=0.25]{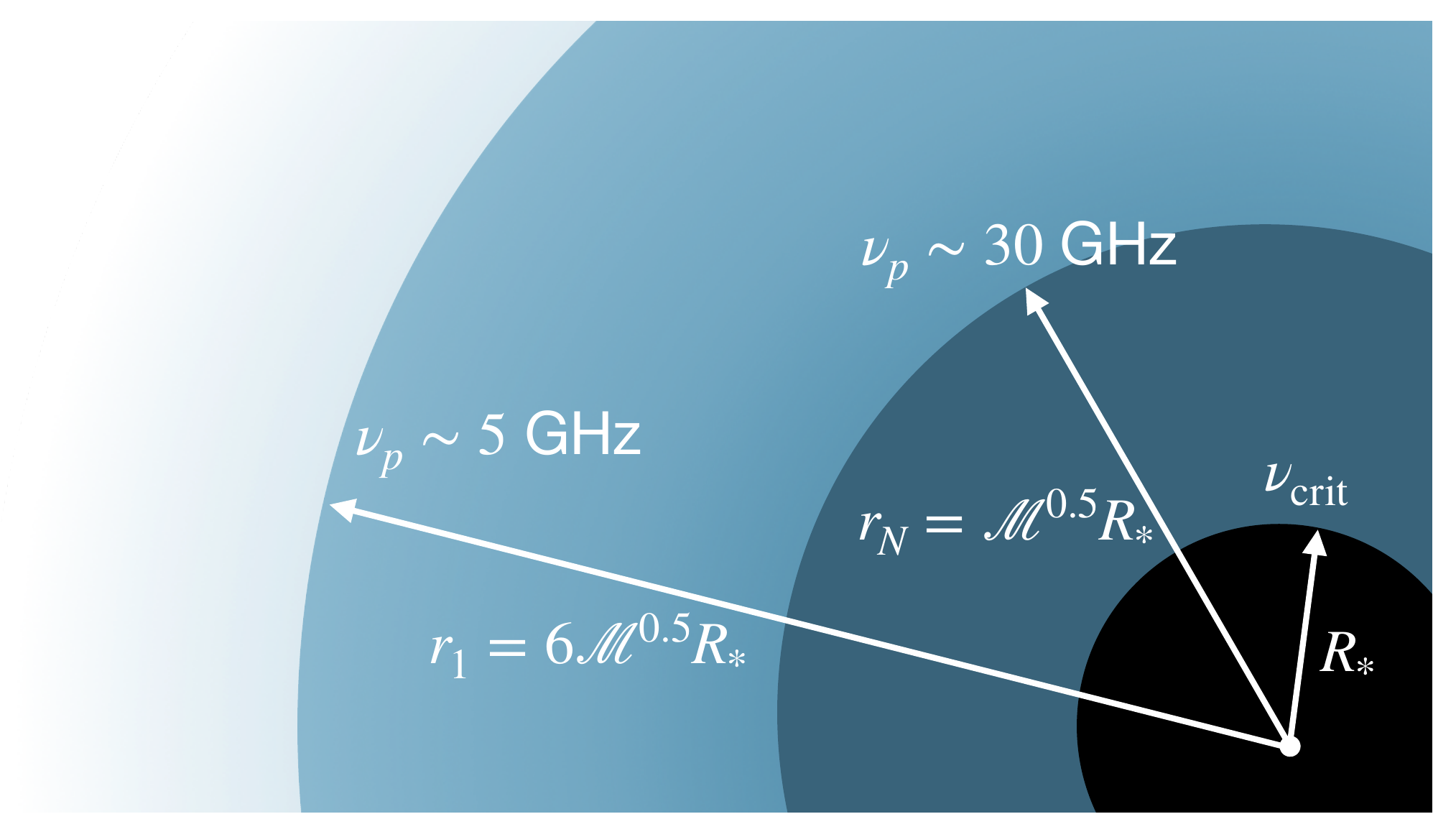}
\caption{A schematic cartoon showing the scales of the magnetospheric region producing the diffractive fringes for the Goldreigh-Julian-like plasma density distribution with arbitrary multiplicity ${\cal M}$. Here $R_*$ is the pulsar radius, $\nu_{p}$ is the plasma frequency, $\nu_{\rm crit}$ is the plasma frequency at the surface.}
\label{f:magn}
\end{figure}

We now estimate the plasma frequency, $\nu_{p}=\omega_{p}/2\pi$, as a function of radius:
\begin{equation}
\nu_{p}(r)\simeq26\ {\cal M}^{1/2}\langle\gamma^{3}\rangle^{-1/2}\left(\frac{r}{R_*}\right)^{-n/2}\textrm{ GHz}.
\label{nup}
\end{equation}
Thus,  if multiplicity is low ${\cal M}\sim1$, the highest observed frequency $\nu_{\rm high}\sim30$~GHz diffracts near the star surface and the lowest frequency $\nu_{\rm low}\sim5$~GHz diffracts at about six times larger distance, as shown in Fig. \ref{f:magn}. 
If the multiplicity is large,  ${\cal M}\gg1$, the corresponding radial scales increase as $r\propto{\cal M}^{1/n}\sim{\cal M}^{0.5}$, assuming that $n=\alpha\approx2$, cf. Eq. (\ref{nup}) for a fixed $\nu_{p}$. We remind that $\gamma\sim1$ on closed field lines is assumed. 

The origin of the low-frequency cutoff $\nu_{\rm low}$ is unknown. For example, it may be associated with the intrinsic  angular extent of the source, or with the outer size of the dipolar magnetosphere. It the latter is true, then it corresponds to the diffraction near the light cylinder radius, $r_{LC}\simeq1.7\times10^{8}$~cm, so that the multiplicity is ${\cal M}\sim10^{3}$, also very reasonable.

The fringe spatial width is estimated to be
\begin{equation}
\Delta_{\rm fringe}\sim d\lambda/a\sim10^{-6}{\cal M}^{-0.5}d\sim 100{\cal M}^{-0.5} \textrm{ AU} ,
\end{equation}
where $d\sim 2$~kpc is the distance to the Crab pulsar. 
The variability timescale due to the relative motion of the pulsar and the Earth, assuming the relative speed of approximately 200~km/s, is
\begin{equation}
\tau\sim\Delta_{\rm fringe}/|v_{\rm Crab}-v_\oplus|\sim 5\times10^7{\cal M}^{-0.5}\textrm{ s }\sim {\cal M}^{-0.5}\textrm{ year}.
\end{equation}
This indicates that the pattern can be very stable on a time-scale of months. 
Shorter variability can be naturally associated with the radiation source instead, both its intrinsic variability and the rotation phase when the emission occurs. The latter is particularly strong because the fringe pattern is very sensitive to the source-screen-observer alignment since the angular fringe width is $(\delta\theta)_{\rm fringe}\sim\lambda/a\sim10^{-6}$. From observations, we know that individual pulses of the HFIP are spread over several degrees in the rotation phase \citep{HE+16a}. Thus, when the radio source is stable in phase, the ``zebra'' pattern is stable as well, whereas when the pulse phase varies, the pattern varies on the pulse-to-pulse case. Interstellar scintillations can also affect the variability. 

At last, we numerically checked that our simplifying assumptions that (i) $n_{\rm refr}=1$ at $r>r_{0}$ and (ii) the ``thin screen approximation'' do not change the qualitative picture. The details can be found in Appendix.

\section{Predictions and conclusion}

In addition to the observational constraints that were mentioned above, the model can be tested on several other predictions in the future.

(1) First, we emphasize that the proportional scaling of the spectral bands is the consequence of the frequency-dependent size of the reflecting region, which plays a role of a diffraction screen. As the observation frequency increases, the corresponding reflection radius $r_{0}(\nu)$ decreases. When it becomes equal to the neutron star radius, the proportional scaling law is modified. Indeed, at higher frequencies, the diffracting screen is the star itself, hence the band spacing should change (e.g., become constant). Thus, we predict the change of the slope of $\Delta \nu$ vs. $\nu$ graph at some frequency $\nu_{\rm crit}$. This break will indicate that $r_{0}(\nu_{\rm crit})=R_*$, as shown in Fig. \ref{f:magn}. The critical frequency 
\begin{equation}
\nu_{\rm crit}=\nu_{p}(R_*)\simeq 26\ {\cal M}^{1/2}\langle\gamma^{3}\rangle^{-1/2}\textrm{ GHz}
\end{equation}
readily follows from Eq. (\ref{nup}). This frequency yields the the plasma multiplicity ${\cal M}$ and the absolute normalization of the plasma density at the neutron star surface in $n_{e}(r)=n_{e}(R_*)\left(r/R_*\right)^{-\alpha}$, namely 
\begin{align}
n_{e}(R_*)&=\nu_{\rm crit}^{2}\pi m_{e}\langle\gamma^{3}\rangle /e^{2}\nonumber\\
&\simeq 1.2\times 10^{14} \langle\gamma^{3}\rangle (\nu_{\rm crit}/100\textrm{ GHz})^{2}\textrm{ cm}^{-3}.
\end{align}

The anticipated $\nu_{\rm crit}$ range is $30\ {\rm GHz}\lesssim \nu_{\rm crit}\lesssim 850\ {\rm GHz}$. The lower bound follows from observations and the model with the minimum reasonable multiplicity ${\cal M}\sim1$. The upper bound assumes the maximum estimated multiplicity of ${\cal M}\sim10^{3}$, when diffraction at 5~GHz occurs near the light cylinder.

Furthermore, we predict that the detection of two neighboring emission bands above $\nu_{\rm crit}$, for instance at $\nu_{i+1}>\nu_i\ge\nu_{\rm crit}$ allows one to determine the fringe order at $\nu_i$. From our model, Eqs. (\ref{a1}), (\ref{a2}) with $a_i=a_{i+1}=2 R_*$, one has $m_i=\nu_i/(\nu_{i+1}-\nu_i)$. Note that the spectral bands are spaced more tightly at $\nu>\nu_{\rm crit}$ since $m_{i}$ is expected to be large, thus $\Delta\nu/\nu=1/m_{i}\ll6\%$. Shall we know somehow the physical distance between the fringes (e.g., from variability) at the Earth location, $\Delta_{{\rm fringe},i}$ (in cm) at frequency $\nu_i$, then $\sin\theta\simeq m_i\Delta_{{\rm fringe},i}/d$, so one can determine the physical pulsar size $R_*\simeq c d/(2\nu_i\Delta_{{\rm fringe},i})$.

(2) Second, the proposed model can be used to perform ``{\em tomography}'' of pulsar magnetospheres. 
We predict that this HFIP properties can also be observed in other pulsars, if their radio and high energy emission are in phase.
This would happen if the radio emission is produced in the outer magnetosphere as opposed to the ``normal'' emission from the polar region. This should be the case for young and millisecond pulsars with high magnetic field strength at the light cylinder. 
The tomography can also be applied to pulsar binaries (e.g., the famous eclipsing double pulsar PSR J0737-3039), where a companion lights up from behind the magnetosphere of a pulsar.

We also want to point to an interesting possible consequence of our model. 
At frequencies where the HFIP appears in the Crab light-curve, two components called the high-frequency components (HFC1 and HFC2) appear as well, but at different rotation phases. They also are much longer than typical emission episodes in the interpulse and main pulse (hundreds of microseconds vs. a few microseconds, respectively). We propose that these high frequency components are the reflections off the magnetosphere of the same source producing the diffracted HFIP. It seems reasonable to believe that the HFIP source emits radiation radially toward the star at various phases. Then, this emission can be reflected from the reflecting region, $r<r_{0}(\nu)$, toward an observer. Since, $r_{0}(\nu)$ decreases with increasing $\nu$, the light path length to the reflection point and then to the observer is longer at higher $\nu$. Thus, the reflected component at higher frequencies should time-lag behind the same component at lower frequencies. Precisely this trend is seen in observations \citep{HE+16a}. Quantitatively, the observed time-lag between 5 and 30~GHz is $\sim10$ degrees of rotation phase, that is about $\delta t\sim 9\times10^{-4}$~s.  Our model predicts $\delta t\sim2\times (r_{0}^{5\textrm{GHz}}-r_{0}^{30\textrm{GHz}})/c\sim2\times (5{\cal M}^{1/2}R_*/c)\sim3\times10^{-4}{\cal M}^{1/2}$~s. The results are fully consistent and yield a reasonable ${\cal M}\sim10$.

To conclude, we propose a model, which explains the peculiar spectral band structure (the ``zebra'' pattern) of the high-frequency interpulse of the Crab pulsar radio emission. The model explains all observed features of the emission, e.g., the proportional separation of the bands (the ``6\% rule''), the persistence and variability of the pattern, the high polarization of the emission and the constancy of the position angle, and the larger dispersion measure variation compared to the main pulse. Certain testable predictions have been made. In addition, the nature of the high-frequency components is addressed. We emphasize that the proposed model would allow to perform ``tomography'' of certain pulsar magnetospheres. Future observations and the detection of the high-frequency turnover in the zebra pattern would yield properties of the magnetospheric plasma such as its density profile, multiplicity, and temperature. These results can be compared with theoretical and numerical models of pulsars. They would thus help one develop more realistic models of pulsar magnetospheres.

\acknowledgements

The author acknowledges useful discussions with Anatoly Spitkovsky, Sasha Philippov, Hayk Hakobyan, Avi Loeb, Ramesh Narayan, Ashley Bransgrove. The author specially thanks Hayk Hakobyan and Sasha Philippov for their very valuable comments on the manuscript. 
This research was partly supported by the National Science Foundation under Grants No. PHY-2010109 and PHY-2409249. 


\bibliography{zebracrab}

\section*{Appendix on the possible radio emission mechanism}

The mechanism involves magnetic reconnection and subsequent violent interaction of plasmoids within the current sheet. These interactions produce `slow plasma modes', which further mode-convert and exit the magnetosphere as electromagnetic ones, see Ref. \citep{P+19} for detail. This model implies that an individual pulse emission is composed of a large number of (blended in time) ``nanoshots'' ---  bright sub-pulses of a few nanosecond duration each --- thus explaining their high temporal variability and broadband spectra. These results are fully consistent with time-resolved spectroscopy of individual emissions of the main pulse, low-frequency interpulse, and several emission components \citep{HE07, HE+16b}. Thus, we can assume that the HFIP source emission also consists of blended nanoshots from reconnection and plasmoid interactions near the LC at $r_{LC}\sim2\times10^{8}$~cm.

\section*{Appendix on mode propagation and absorption}

Electromagnetic waves, propagating in magnetized plasma quasi-perpendicular to the magnetic field have two distinct polarizations, the `ordinary' and `extraordinary' modes (O- and X-modes, for short). These two polarizations have different propagation properties. The O-mode has the wave electric field along the ambient field, so it ``feels'' plasma but not the field. Thus, its cutoff frequency is the relativistic plasma frequency, $\omega_{p}=\sqrt{4 \pi e^{2} n_{e}/\langle\gamma^{3}\rangle m_{e}}$. Plasma can be highly relativistic, $\gamma\gg1$, on open field lines and sub-relativistic on closed field lines, because of cooling of the trapped plasma. As is said above, we assume that diffraction happens in the inner magnetosphere on closed field lines, so one expects $\gamma\sim 1$. The O-mode cannot propagate in a region, where $\omega<\omega_p$, but instead reflects off it. In contrast, the X-mode does ``feel'' magnetic field, so it has a cutoff and a resonance near the cyclotron frequency, $\omega_{B}=eB/\gamma m_e c$. Cyclotron resonant absorption strongly limits the wave propagation through a magnetosphere \citep{BS76, M+82}. The optical depth in the Crab magnetosphere is $\tau_{\rm abs}\sim 0.1{\cal M}(\nu/{\rm 30~GHz})^{-1/3}$ \citep[cf. Eq. (8) in Ref.][]{RG05}. Thus, X-mode is strongly absorbed ($\tau_{\rm abs}\gtrsim1$) in a plasma with multiplicity ${\cal M}\gtrsim10$. In the Grab, ${\cal M}\sim10^4-10^5$ on open field lines and somewhat lower on closed field lines \citep{PK22}. It is therefore reasonable to believe that only O-mode can propagate through the Crab magnetosphere of closed field lines without severe attenuation.

\section*{Appendix on numerical solution}

\begin{figure*}
\includegraphics[scale=0.9]{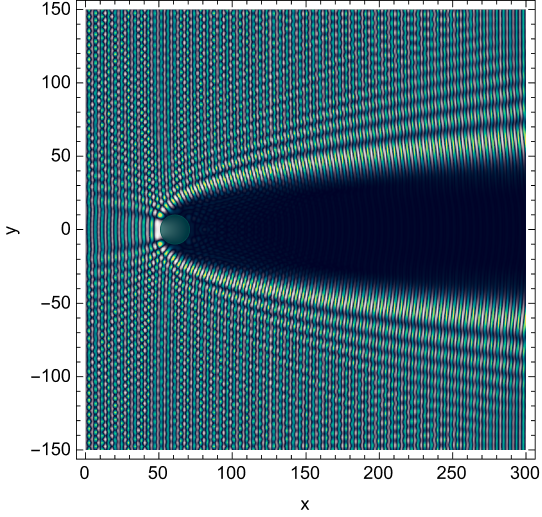} \hspace{10pt}
\includegraphics[scale=0.9]{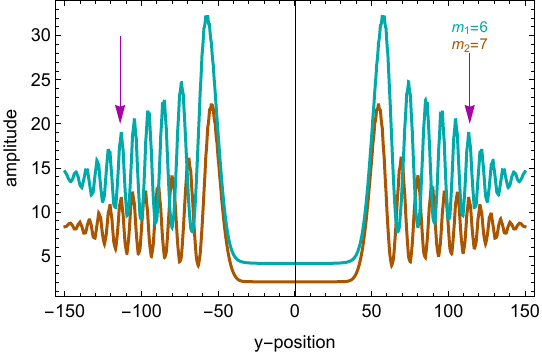}
\caption{Modeling wave diffraction off a circular reflecting region with radially varying index of refraction outside of it.  Both panels use arbitrary units for all variables. {\it Upper panel:} The snapshot of the wave equation solution. A plane wave is produced at the left boundary. The index of refraction is given by Eq. (\ref{nrefr}). The reflecting region, $r<r_{0}$, where $n_{\rm refr}^{2}\le0$ is shown as a dark  disk. {\it Lower panel:} The diffraction patterns for two frequencies, $\omega_{1}=0.9$ and $\omega_{2}=1$ (cyan and brown curves, respectively), are obtained near the right end of the box. Note the overlap of the corresponding fringes $m_{1}=6$ and $m_{2}=7$ illustrates how a ``zebra'' spectral band pattern emerges. The curves are shifted vertically, for clarity. }
\label{f:waveeq}
\end{figure*}

Here we check that our simplifying assumptions that (i) $n_{\rm refr}=1$ at $r>r_{0}$ and (ii) the ``thin screen approximation'' do not change the qualitative picture. For this, we solve the wave equation in a two-dimensional domain with the index of refraction of the O-mode given by 
\begin{equation}
n_{\rm refr}=\left(1-(r/r_0)^{-2}\right)^{1/2}
\label{nrefr}
\end{equation}
and the reflection radius $r_{0}=a(\omega)/2$. 
This equation readily follows from Eqs. (\ref{nOmode}), (\ref{ne}) with $\alpha=2$. Note, the wave cannot propagate into and is reflected from the region, where $n_{\rm refr}^{2}\le0$, i.e., at $r<r_{0}$. The results are shown in Fig. \ref{f:waveeq}. All units are arbitrary. The wave equation solution is performed in a box with reflecting boundary conditions. A plane wave is injected at the left wall, as is shown in the first panel. The reflecting region, $r<r_{0}$,  is marked by the dark cyan disk centered at coordinates $(60,0)$. Here we use $r_{0}=11$ and $\omega=0.9$, which corresponds to the wavelength $\lambda\approx3.9$. The diffraction patterns shown in the second panel are taken near the right wall before the waves reflected from the walls have reached it. The results for two frequencies, $\omega_{1,2}=0.9, 1$, are shown with cyan and brown curves respectively, with the corresponding $r_{0}(\omega_{1,2})=11, 10$. In both cases, the diffraction pattern is clearly seen. The presence of a radially-dependent index of refraction, $n_{\rm refr}(r)$, and the circular shape of the reflecting region do not prevent the pattern from formation. The coincidence of two maxima at the same location (marked by arrows) indicates that two bands, $\omega_{1},\omega_{2}$, would be seen there.

\end{document}